\documentclass{article}
\usepackage[T1]{fontenc}
\usepackage[utf8]{inputenc}
\usepackage{booktabs}
\usepackage{multirow}
\usepackage{makecell}
\usepackage{ismir}
\usepackage{amsmath,cite,url,amssymb}
\usepackage{graphicx}
\usepackage{subcaption}
\usepackage{color}
\usepackage[table]{xcolor} 
\usepackage{pifont}
\newcommand{\cmark}{\ding{51}}

\title{CultureMERT: Continual Pre-Training for Cross-Cultural Music Representation Learning}

\multauthor
{Angelos-Nikolaos Kanatas$^{1,2}$ \hspace{1cm} Charilaos Papaioannou$^{1,3,4}$ \hspace{1cm} Alexandros Potamianos$^{1,4}$} {
 $^{1}$ School of ECE, National Technical University of Athens, Greece\\
 $^{2}$ Institute for Language and Speech Processing, Athena Research Center, Greece\\
 $^{3}$ Centre for Digital Music, Queen Mary University of London, UK\\
 $^{4}$ Archimedes, Athena Research Center, Greece\\
{\tt\small el19169@mail.ntua.gr, cpapaioan@mail.ntua.gr, potam@central.ntua.gr}
}

\def\authorname{A.-N. Kanatas, C. Papaioannou, and A. Potamianos}

\usepackage[bookmarks=false,pdfauthor={\authorname},pdfsubject={\pdfsubject},hidelinks]{hyperref}

\begin{document}

\maketitle

\begin{abstract}

\noindent
Recent advances in music foundation models have improved audio representation learning, yet their effectiveness across diverse musical traditions remains limited. We introduce \texttt{CultureMERT-95M}, a multi-culturally adapted foundation model developed to enhance cross-cultural music representation learning and understanding. To achieve this, we propose a two-stage continual pre-training strategy that integrates learning rate re-warming and re-decaying, enabling stable adaptation even with limited computational resources. Training on a 650-hour multi-cultural data mix, comprising Greek, Turkish, and Indian music traditions, results in an average improvement of 4.9\% in ROC-AUC and AP across diverse non-Western music auto-tagging tasks, surpassing prior state-of-the-art, with minimal forgetting on Western-centric benchmarks. We further investigate task arithmetic, an alternative approach to multi-cultural adaptation that merges single-culture adapted models in the weight space. Task arithmetic performs on par with our multi-culturally trained model on non-Western auto-tagging tasks and shows no regression on Western datasets.
Cross-cultural evaluation reveals that single-culture models transfer with varying effectiveness across musical traditions, whereas the multi-culturally adapted model achieves the best overall performance. To support research on world music representation learning, we publicly release \texttt{CultureMERT-95M}\footnote{\href{https://huggingface.co/ntua-slp/CultureMERT-95M}{https://huggingface.co/ntua-slp/CultureMERT-95M}} and \texttt{CultureMERT-TA-95M}\footnote{\href{https://huggingface.co/ntua-slp/CultureMERT-TA-95M}{https://huggingface.co/ntua-slp/CultureMERT-TA-95M}}, fostering the development of more culturally aware music foundation models.

\end{abstract}

\section{Introduction}\label{sec:introduction}

Foundation models have recently emerged in the music domain \cite{DBLP:conf/iclr/LiYZMCYXLRBGDLC24, DBLP:conf/icassp/WonHL24, DBLP:journals/corr/abs-2005-00341, DBLP:journals/corr/abs-2408-14340, DBLP:journals/corr/abs-2409-09601}, offering powerful general-purpose representations learned from large-scale audio data. These models capture broad musical characteristics and have demonstrated state-of-the-art performance across a range of music understanding tasks, reducing the need for task-specific training. By leveraging self-supervised learning (SSL) on large amounts of unlabelled music data, foundation models address data scarcity, reduce annotation costs, and improve generalization in music information retrieval (MIR) \cite{DBLP:journals/corr/abs-2408-14340}.

Despite these advances, most existing foundation models for music have been trained primarily on Western-centric datasets, limiting their ability to represent diverse musical styles \cite{gomez_computational_2013, DBLP:journals/corr/abs-2502-07328}. Many musical traditions, including Turkish, Indian, and Greek traditional music, feature unique melodic structures, modal or tonal systems, and rhythmic patterns that are not adequately captured by these models \cite{DBLP:journals/sigpro/LidySCGRKK10, DBLP:journals/tismir/PlajaRoglansNPSM23, DBLP:conf/ismir/KoduriMSS11}.
Failing to model such culture-specific stylistic elements not only narrows the applicability of music foundation models, for example, in region-specific recommendation systems \cite{DBLP:journals/tors/FerraroFDB24} or cultural heritage preservation, but also overlooks rich, culturally specific knowledge crucial for advancing MIR research \cite{DBLP:journals/corr/abs-2408-14340}.
Accordingly, there is an urgent need to develop more inclusive and culturally aware computational models \cite{DBLP:journals/corr/abs-2406-03930, DBLP:journals/tismir/HolzapfelSC18}, capable of generalizing beyond Western-centric traditions and adapting effectively to diverse underrepresented musical cultures.

One promising avenue for addressing these challenges is continual pre-training (CPT), which has emerged as an effective and increasingly popular approach in large language models (LLMs) \cite{DBLP:journals/tmlr/IbrahimTGRABLR24, DBLP:journals/corr/abs-2402-17733, DBLP:journals/corr/abs-2407-20743, DBLP:conf/tsd/GogoulouLBN24, DBLP:conf/acl/GururanganMSLBD20, DBLP:journals/corr/abs-2407-07263, DBLP:journals/corr/abs-2404-17790, DBLP:journals/corr/abs-2308-04014, DBLP:journals/corr/abs-2404-16789} and multimodal learning \cite{DBLP:conf/nips/UdandaraoRDPCVH24}. By enabling models to incrementally adapt to new domains, tasks, or languages, CPT avoids the need for full re-training, which is often impractical and computationally expensive \cite{DBLP:conf/acl/GururanganMSLBD20, DBLP:journals/tmlr/IbrahimTGRABLR24, DBLP:journals/corr/abs-2407-07263, DBLP:journals/corr/abs-2404-16789, DBLP:journals/ipm/NowakowskiPMN23}. Notably, it has been shown to match, or even surpass, training from scratch in some cases \cite{DBLP:journals/corr/abs-2404-17790, DBLP:journals/corr/abs-2308-04014}, while also converging faster \cite{DBLP:conf/emnlp/ZhengPXQ0Z24} and mitigating catastrophic forgetting \cite{DBLP:journals/nn/CossuCPLTB24}.
CPT has also gained traction in the audio domain, with recent work demonstrating its effectiveness in adapting pre-trained speech models to both high- and low-resource languages \cite{DBLP:journals/ipm/NowakowskiPMN23, DBLP:journals/corr/abs-2207-00659, DBLP:journals/taslp/ZhuCWHZY24, DBLP:conf/sigtype/SanPAHKAJ24, DBLP:journals/corr/abs-2409-14494}.

Additionally, model merging \cite{DBLP:conf/icml/WortsmanIGRLMNF22, DBLP:journals/corr/abs-2408-07666, DBLP:journals/corr/abs-2412-12153, DBLP:conf/nips/IlharcoWGSHKFS22} has proven to be a simple yet effective technique for adapting pre-trained models across multiple domains by combining domain-specific parameters in weight space, without requiring additional training \cite{DBLP:conf/iclr/StoicaBBRHH24} or access to the original training data \cite{DBLP:conf/iclr/Jin0P023}. A notable method within this paradigm is task arithmetic (TA) \cite{DBLP:conf/iclr/IlharcoRWSHF23}, which constructs \textit{task vectors} by computing the difference between the parameters of an adapted model and its pre-trained counterpart, thereby encoding domain-specific knowledge. These task vectors can then be integrated into the pre-trained model via algebraic operations in Euclidean space to create a unified model from multiple independently adapted models.

While both continual pre-training and task arithmetic have been widely explored in other domains, their application to MIR remains largely unexplored. We bridge this gap by leveraging these techniques to adapt the \texttt{MERT-v1-95M} music foundation model \cite{DBLP:conf/iclr/LiYZMCYXLRBGDLC24}, originally trained on 1K hours of predominantly Western music \cite{DBLP:conf/iclr/LiYZMCYXLRBGDLC24, DBLP:conf/icassp/LiMWKWCXB024}, to diverse musical cultures from the Eastern Mediterranean and the Indian subcontinent, while preserving performance on "Western"-centric benchmarks.

We summarize our main contributions as follows:
\begin{enumerate}

    \item To the best of our knowledge, this is the first study to explore \textbf{continual pre-training} and \textbf{task arithmetic} for cross-cultural adaptation in MIR, demonstrating their effectiveness in music audio representation learning.

    \item We propose a \textbf{two-stage CPT strategy} that stabilizes training, mitigates catastrophic forgetting, and facilitates effective adaptation under constrained computational resources.

    \item Our multi-cultural model, \texttt{CultureMERT}, outperforms the original \texttt{MERT-v1} by an average of \textbf{4.9\%} across ROC-AUC and AP on culturally diverse non-Western music tagging tasks, while exhibiting minimal forgetting on Western benchmarks.

    \item Our culturally adapted models \textbf{surpass previous state-of-the-art} results across all evaluated non-Western music tagging tasks.

    \item We analyze \textbf{cross-cultural transferability}, showing that single-culture adaptations exhibit varying degrees of transfer across cultural domains.

\end{enumerate}

To support reproducibility and further research in cross-cultural music representation learning, we publicly release \texttt{CultureMERT-95M}, along with the task arithmetic variant, \texttt{CultureMERT-TA-95M}.

\section{Datasets}\label{sec:data}

For our experiments, we use a diverse set of music datasets spanning both Western and non-Western traditions. Specifically, we adopt the MagnaTagATune (MTAT) \cite{DBLP:conf/ismir/LawWMBD09} and FMA-medium \cite{DBLP:conf/ismir/DefferrardBVB17} datasets to represent "Western"\footnote{We use the term “Western” to refer to music styles predominantly rooted in Western cultures, including pop, rock, and Western classical.} music. For "non-Western" traditions, we incorporate the Lyra corpus \cite{DBLP:conf/ismir/PapaioannouVGKP22}, featuring Greek traditional and folk music, along with three collections from the CompMusic Corpora\footnote{\href{https://compmusic.upf.edu/corpora}{https://compmusic.upf.edu/corpora}} \cite{DBLP:conf/semanticaudio/Serra14}: Turkish-makam \cite{DBLP:conf/jcdl/UyarASBS14, DBLP:phd/es/Senturk17}, which, together with Lyra, represent music of the Eastern Mediterranean; and Hindustani and Carnatic music \cite{DBLP:conf/icmc/Srinivasamurthy14}, representing North and South Indian classical traditions, respectively.

We assess our models on both Western and non-Western music tagging tasks for cross-cultural evaluation, using standard multi-label classification metrics, including the area under the receiver operating characteristic curve (ROC-AUC) and average precision (AP). Following \cite{DBLP:conf/ismir/PapaioannouBP23, papaioannou_lc-protonets_2025}, we utilize the top-k tags relevant to each dataset: 50 tags for MTAT (spanning \textit{genre}, \textit{instruments}, and \textit{mood}), 20 hierarchical \textit{genre} tags for FMA-medium, 30 tags for Turkish-makam (covering \textit{makam}, \textit{usul}, and \textit{instruments}), 20 tags for Hindustani and Carnatic (primarily reflecting \textit{raga}, \textit{tala}, \textit{instruments}, and \textit{forms}), and 30 tags for Lyra (related to \textit{genre}, \textit{place}, and \textit{instruments}).

All audio is resampled to 24 kHz, and we adopt the same data splits as \cite{DBLP:conf/ismir/PapaioannouBP23}. To prepare our data for continual pre-training, we extract 30-second segments from each training split of the non-Western datasets. Given the varying dataset sizes, we balance the pre-training duration across cultures to ensure proportional representation by extracting 200 hours each from the Turkish-makam, Carnatic, and Hindustani datasets, and 50 hours from Lyra due to its smaller size. Additionally, we combine these subsets to construct a unified 650-hour dataset integrating all four traditions for multi-cultural continual pre-training.

\section{Method}\label{sec:method}

The overall framework of our approach is illustrated in Figure~\ref{fig:method_figure}, which depicts the two-stage continual pre-training strategy for \texttt{CultureMERT}. In this section, we first review the architecture and pre-training objective of MERT, and then present our CPT strategy for cultural adaptation. Finally, we investigate task arithmetic, an alternative approach to multi-cultural adaptation that merges culturally specialized models in weight space to construct a unified multi-cultural model, \texttt{CultureMERT-TA}.

\subsection{MERT Pre-Training Objective}\label{subsec:mert_objective}

Our continual pre-training objective follows the self-supervised masked language modeling (MLM) objective of \texttt{MERT\textsuperscript{RVQ-VAE}}, where two teacher models provide the pseudo-labels: (i) an \textit{acoustic teacher}, the EnCodec model \cite{DBLP:journals/tmlr/DefossezCSA23}, which discretizes audio into tokens from $K=8$ residual vector quantization (RVQ) codebooks, each containing $C=1024$ codewords, and (ii) a \textit{musical teacher}, based on constant-Q transform (CQT) spectrogram reconstruction, encoding pitch and harmonic structure.

\texttt{MERT-v1-95M} follows the HuBERT architecture \cite{DBLP:journals/taslp/HsuBTLSM21}, comprising a CNN-based feature extractor that encodes raw 24 kHz waveforms into 75 Hz frame-level representations, followed by a 12-layer Transformer encoder, producing 768-dimensional contextual embeddings. During training, a subset of frame embeddings is masked, and the model is optimized using a multi-task learning (MTL) objective, combining masked acoustic token prediction and spectrogram reconstruction.
The overall training objective is:
\begin{equation}
\mathcal{L} = \alpha \mathcal{L}_{\text{RVQ}} + \mathcal{L}_{\text{CQT}},
\end{equation}
where the acoustic MLM loss $\mathcal{L}_{\text{RVQ}}$ encourages the model to predict masked RVQ-VAE tokens from $K$ codebooks, using a noise-contrastive estimation (NCE) loss:
\begin{equation}
\mathcal{L}_{\text{RVQ}} = \sum_{k=1}^{K} \sum_{t \in M} \log p_{\theta}(c_{t, k} | \boldsymbol{x}'_t),
\end{equation}
with \(M\) denoting the set of masked time frames, \(c_{t, k}\) the \textit{ground-truth} discrete codeword from the \(k\)-th codebook at time frame \(t\) extracted via the EnCodec tokenizer, and \(p_{\theta}\) the model’s predicted token distribution:
\begin{equation}
p_{\theta}(c | \boldsymbol{x}'_t) = \frac{\exp(\text{sim}(T(\boldsymbol{o}_t), \boldsymbol{e}_c)/\tau)}
{\sum_{c'=1}^{C} \exp(\text{sim}(T(\boldsymbol{o}_t), \boldsymbol{e}_{c'})/\tau)}.
\end{equation}
Here, $\boldsymbol{x}'_t$ is the masked input feature, $\boldsymbol{o}_t$ is the model’s output representation, $T(\boldsymbol{o}_t)$ projects it to the codeword embedding space, $\boldsymbol{e}_c$ is the embedding of codeword $c \in \mathcal{C}_k$, where $k \in \{1, \dots, K\}$, $\text{sim}(\cdot, \cdot)$ denotes cosine similarity, and $\tau=0.1$ is a temperature scaling parameter.

The CQT reconstruction loss \(\mathcal{L}_{\text{CQT}}\) minimizes the mean squared error (MSE) between the model's predicted \(\hat{\mathbf{z}}_{\text{CQT}, t}\) and ground-truth \(\mathbf{z}_{\text{CQT}, t}\) frame-level CQT features:
\begin{equation}
\mathcal{L}_{\text{CQT}} = \sum_{t \in M} \left\| \mathbf{z}_{\text{CQT}, t} - \hat{\mathbf{z}}_{\text{CQT}, t} \right\|_2^2.
\end{equation}

\subsection{Two-Stage Continual Pre-Training Strategy}\label{subsec:twostage_cpt_method}

To adapt the MERT foundation model to diverse musical traditions, we employ continual pre-training, which extends the training of a pre-trained model on new data, aiming to adapt it to a shifted domain or task while retaining prior knowledge, without re-training from scratch. In our case, this involves continually pre-training the \texttt{MERT-v1-95M} model, using the same pre-training objective, on culturally diverse data that introduce a significant distribution shift, as it was initially trained on predominantly Western music \cite{DBLP:conf/iclr/LiYZMCYXLRBGDLC24, DBLP:conf/icassp/LiMWKWCXB024}. 
Given this shift, naively continuing to train the model, i.e., adapting all parameters at once without resetting the learning rate, can lead to catastrophic forgetting \cite{DBLP:journals/corr/KirkpatrickPRVD16} and poor adaptation \cite{DBLP:journals/tmlr/IbrahimTGRABLR24}, as confirmed by our preliminary experiments (see Table~\ref{tab:cpt_ablations_minimal}). To address this, we propose a \textbf{two-stage} strategy that stabilizes training through: (i) learning rate re-warming and re-decaying \cite{DBLP:journals/tmlr/IbrahimTGRABLR24, DBLP:conf/nips/UdandaraoRDPCVH24, DBLP:journals/corr/abs-2308-04014, DBLP:journals/corr/abs-2407-07263, DBLP:journals/corr/abs-2404-06395}, and (ii) staged adaptation.

\textbf{Staged Adaptation}
In our preliminary experiments, we observed an initial performance drop during CPT, followed by a slow recovery phase, a phenomenon known as the \textit{stability gap} \cite{DBLP:journals/corr/abs-2406-14833, DBLP:conf/iclr/LangeVT23, DBLP:conf/nips/UdandaraoRDPCVH24}. This instability arises due to the abrupt adaptation of model parameters to a substantially shifted data distribution, which can temporarily degrade previously learned representations before stabilizing. To mitigate this, rather than full-parameter adaptation on the entire dataset in a single epoch, which induces a large plasticity gradient for a long period \cite{DBLP:conf/iclr/LangeVT23}, we split training into two stages to reduce instability and ensure smoother adaptation, as illustrated in Figure \ref{fig:method_figure}:

\begin{figure}[t]
  \hspace{-2.3em}
  \includegraphics[width=0.56\textwidth]{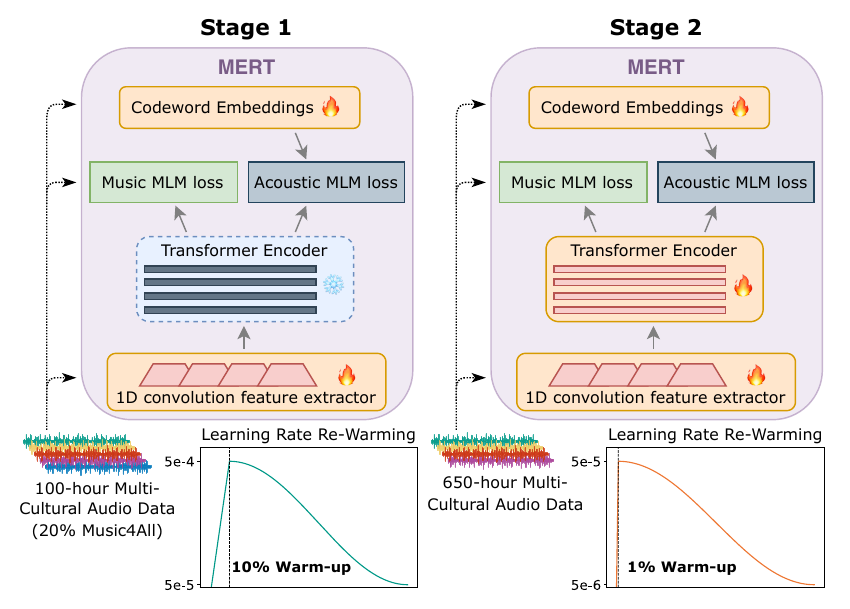}
  \caption{\textbf{Two-Stage Continual Pre-Training Strategy for} \texttt{CultureMERT}. In Stage 1, a subset of parameters is trained on 100h of multi-cultural data with 20\% Western music for stabilization. In Stage 2, all parameters are unfrozen and trained on the full 650h dataset. Learning rate re-warming and re-decaying is applied in both stages.}
  \label{fig:method_figure}
\end{figure}

\textbf{Stage 1}
\textit{Stabilization Phase}: We first train on a smaller data subset \cite{DBLP:journals/corr/abs-2406-14833}, updating only the CNN-based feature extractor and the codeword embedding layer while keeping the Transformer encoder frozen. To reduce the distribution gap and mitigate forgetting \cite{DBLP:journals/corr/abs-2407-07263, DBLP:journals/taslp/ZhuCWHZY24, DBLP:conf/emnlp/ZhengPXQ0Z24}, we incorporate a fraction of Music4All data \cite{DBLP:conf/iwssip/SantanaPDCMCFD20}, which is primarily of Western origin, into the pre-training mix, accounting for 20\% of the total training data (\textit{Western replay}).

\textbf{Stage 2}
\textit{Full Adaptation}: We unfreeze the Transformer encoder and continue training on the full dataset.

\begin{table}[ht]
\centering
\renewcommand{\arraystretch}{1.3}
\setlength{\tabcolsep}{2.1pt}
\footnotesize
\begin{tabular}{l|c|cc}
\toprule
\textbf{CPT Strategy} & \textbf{Western Replay} & \textbf{Turkish-makam} & \textbf{MTAT} \\
\midrule
\midrule
\rowcolor{gray!20} \multicolumn{1}{l|}{\hspace{1em}MERT-v1 (Baseline)} & - & 83.2 & \textbf{89.6} \\
\midrule
Single-stage & \cmark & 83.8 & 86.0 \\
Single-stage (no re-warm) & \cmark & 83.0 & 87.5 \\
\midrule
Two-stage (\textit{Ours}) & Stage 1 & \textbf{89.6} & 89.2 \\
Two-stage (\textit{Ours}) & Both stages & 88.6 & 89.4 \\
\bottomrule
\end{tabular}
\caption{\textbf{CPT Strategy Comparison.} ROC-AUC scores on Turkish-makam and MTAT datasets. Two-stage CPT outperforms single-stage adaptation, with \textit{Western replay} limited to Stage 1 yielding the best trade-off between cultural adaptation and knowledge retention.}
\label{tab:cpt_ablations_minimal}
\end{table}

This two-stage approach is particularly motivated by computational constraints, specifically the batch size mismatch between pre-training and adaptation. \texttt{MERT-v1-95M} was originally trained with batch sizes of 1.5 hours per step, whereas we use a significantly smaller effective batch size of 160 seconds per step due to memory limitations. Training with this reduced batch size directly on the entire dataset with full-parameter adaptation resulted in unstable training and frequent crashes, degrading performance on both Western and non-Western benchmarks.
By structuring adaptation in two stages, we strike to balance \textit{plasticity} (adaptation to non-Western traditions) and \textit{stability} (retaining knowledge on Western datasets), a challenge known as the \textit{stability-plasticity dilemma} \cite{DBLP:journals/nn/ParisiKPKW19, DBLP:conf/cvpr/KimH23}.
Intuitively, the initial \textit{stabilization phase} allows lower-level acoustic representations, captured by the CNN-based feature extractor and the codeword embeddings, to adapt first and calibrate to the shifted distribution before updating high-level Transformer representations.

\textbf{Learning Rate Re-Warming}
To further improve adaptation stability, we apply learning rate re-warming and re-decaying in both stages. Prior work has shown that resetting the learning rate schedule, i.e., \textit{re-warming} the model, during continual pre-training is crucial for preventing poor convergence and mitigating catastrophic forgetting
\cite{DBLP:journals/tmlr/IbrahimTGRABLR24, DBLP:conf/nips/UdandaraoRDPCVH24, DBLP:journals/corr/abs-2308-04014, DBLP:journals/corr/abs-2407-07263, DBLP:journals/corr/abs-2404-06395}. In Stage 1, we adopt a moderately aggressive warm-up and decay schedule to encourage early adaptation of low-level representations. In Stage 2, a less aggressive schedule balances plasticity and stability during full-model training, reducing also training instabilities.

Following this two-stage CPT strategy, we develop two types of culturally adapted models: (i) a \textbf{multi-culturally adapted model}, \texttt{CultureMERT}, trained on a culturally diverse mix spanning all four non-Western musical traditions; and (ii) \textbf{single-culture adapted models}, each continually pre-trained on data from a single tradition, resulting in \texttt{MakamMERT}, \texttt{HindustaniMERT}, \texttt{CarnaticMERT}, and \texttt{LyraMERT}. 

\subsection{Task Arithmetic for Cross-Cultural Adaptation}\label{subsec:ta_method}

As an alternative to continual pre-training on multi-cultural data, we explore task arithmetic \cite{DBLP:conf/iclr/IlharcoRWSHF23}, a model merging method that combines culturally specialized models in weight space to construct a unified multi-cultural model. Task arithmetic operates by algebraically merging model parameters through task vector addition and negation.

In our setting, we obtain \textit{task vectors} by computing the element-wise difference between the parameters of the single-culture continually pre-trained models and those of the \texttt{MERT-v1} model. Formally, given the pre-trained base model with parameters $\theta_{\text{pre}}$ and a continually pre-trained model $\theta_i$ adapted to a cultural dataset $\mathcal{D}_i$, the task vector for culture $i$ is given by $\tau_i = \theta_i - \theta_{\text{pre}}$, capturing the parameter shift induced by culture-specific adaptation.

For multi-cultural adaptation, we construct a unified model $\theta_{\text{merged}}$ by merging $N$ single-culture adapted models via task arithmetic, summing their respective task vectors $\tau_i$ with corresponding scaling factors $\lambda_i$: 
\begin{equation} 
    \theta_{\text{merged}} = \theta_{\text{pre}} + \sum_{i=1}^{N} \lambda_i \tau_i, 
\label{eq:task_arithmetic} 
\end{equation}
where $\lambda_i \in \mathbb{R}$ are scalar hyperparameters that control the contribution of each task vector.
Prior work typically uses a single scaling factor $\lambda$ for all task vectors, i.e., $\lambda_i = \lambda$, $\forall i$. 
In the special case where $\lambda = 1/N$, Equation~\ref{eq:task_arithmetic} simplifies to \textit{weight averaging} \cite{DBLP:journals/corr/abs-2412-12153, DBLP:conf/nips/IlharcoWGSHKFS22, DBLP:conf/icml/WortsmanIGRLMNF22}, in which the adapted models are merged by directly averaging their parameters.

\section{Experiments}\label{sec:experiments}

\subsection{Implementation Details}\label{subsec:implementation}

In all continual pre-training setups, we initialize our models from the publicly available \texttt{MERT-v1-95M}\footnote{\href{https://huggingface.co/m-a-p/MERT-v1-95M}{https://huggingface.co/m-a-p/MERT-v1-95M}} pre-trained checkpoint. Training was conducted using the \textsc{fairseq}\footnote{\href{https://github.com/facebookresearch/fairseq}{https://github.com/facebookresearch/fairseq}} framework on a single NVIDIA GeForce GTX TITAN X GPU with 12 GB of memory. All models were trained with half-precision (FP16), using 5-second audio segments as input context, randomly cropped from the extracted 30-second pre-training audio data. The weight of the acoustic loss in the pre-training objective is set to $\alpha = 10.0$. The EnCodec neural audio codec (NAC) model \cite{DBLP:journals/tmlr/DefossezCSA23}, which tokenizes audio into discrete codewords, remains frozen during continual pre-training, as in \cite{DBLP:conf/iclr/LiYZMCYXLRBGDLC24}. To enhance representation robustness, we apply \textit{in-batch noise mixture augmentation} with a mixup probability of $0.5$, and use pre-layer normalization (Pre-LN) \cite{DBLP:conf/icml/XiongYHZZXZLWL20} for training stability, following \cite{DBLP:conf/iclr/LiYZMCYXLRBGDLC24}. Other training settings mirror those of the \texttt{MERT-v1-95M} setup.

\subsection{Probing-Based Evaluation}\label{subsec:evaluation}

Following \cite{DBLP:conf/ismir/CastellonDL21, DBLP:conf/iclr/LiYZMCYXLRBGDLC24, DBLP:conf/icassp/WonHL24}, we adopt a probing-based evaluation rather than fine-tuning, keeping the pre-trained models frozen as deep feature extractors while training only a shallow multilayer perceptron (MLP) with a single 512-dimensional hidden layer for sequence-level tasks. Our evaluation follows the MARBLE protocol \cite{DBLP:conf/nips/YuanMLZCYZLHTDW23} under constrained settings, and we apply it to both Western and non-Western music tagging tasks for cross-cultural evaluation. To process long-duration audio files, we segment them into 30-second chunks using a sliding window approach and aggregate the chunk-level predictions by averaging to obtain the final prediction for the entire audio file. For Turkish-makam, Hindustani, and Carnatic tasks, we apply a maximum duration cut as in \cite{DBLP:conf/ismir/PapaioannouBP23} to ensure comparability with prior state-of-the-art results.

\subsection{Continual Pre-Training Settings}\label{subsec:cpt_settings}

\begin{table*}[t]
  \renewcommand{\arraystretch}{1.2}
  \setlength{\tabcolsep}{2pt}
  \centering
  \footnotesize
  \begin{tabular}{l|cc|cc|cc|cc|cc|cc|c}
    \toprule
    \textbf{Dataset} 
    & \multicolumn{2}{c|}{\textbf{Turkish-makam}} 
    & \multicolumn{2}{c|}{\textbf{Hindustani}} 
    & \multicolumn{2}{c|}{\textbf{Carnatic}} 
    & \multicolumn{2}{c|}{\textbf{Lyra}}
    & \multicolumn{2}{c|}{\textbf{FMA-medium}} 
    & \multicolumn{2}{c|}{\textbf{MagnaTagATune}} 
    & \multirow{2}{*}{\textbf{Avg.}}  \\
    \cmidrule(lr){2-3} \cmidrule(lr){4-5} \cmidrule(lr){6-7} \cmidrule(lr){8-9} \cmidrule(lr){10-11} \cmidrule(lr){12-13}
    \textbf{Metrics} 
    & \textbf{ROC} & \textbf{AP} 
    & \textbf{ROC} & \textbf{AP} 
    & \textbf{ROC} & \textbf{AP} 
    & \textbf{ROC} & \textbf{AP} 
    & \textbf{ROC} & \textbf{AP} 
    & \textbf{ROC} & \textbf{AP} 
    & \\
    \midrule
    \midrule
    MERT-v1
    & 83.2$_{0.08}$ & 53.3$_{0.12}$ 
    & 82.4$_{0.04}$ & 52.9$_{0.19}$ 
    & 74.9$_{0.05}$ & 39.7$_{0.15}$ 
    & 85.7$_{0.10}$ & 56.5$_{0.18}$ 
    & 90.7$_{0.04}$ & 48.1$_{0.11}$ 
    & 89.6$_{0.07}$ & 35.9$_{0.15}$
    & 66.1 \\
    \midrule
    MakamMERT
    & 88.7$_{0.11}$ & 58.8$_{0.22}$ 
    & 84.5$_{0.16}$ & 57.8$_{0.18}$ 
    & 77.6$_{0.14}$ & 42.7$_{0.16}$ 
    & 84.6$_{0.12}$ & 53.2$_{0.17}$ 
    & 90.3$_{0.12}$ & 47.1$_{0.16}$ 
    & 89.0$_{0.07}$ & 35.6$_{0.12}$
    & 67.5 \\
    CarnaticMERT
    & 88.4$_{0.06}$ & 58.4$_{0.16}$ 
    & 87.0$_{0.06}$ & 60.2$_{0.14}$ 
    & 78.8$_{0.13}$ & \textbf{44.0}$_{0.17}$ 
    & 85.4$_{0.11}$ & 55.8$_{0.16}$ 
    & 90.2$_{0.10}$ & 46.7$_{0.09}$ 
    & 89.2$_{0.10}$ & 35.3$_{0.11}$
    & 68.3 \\
    HindustaniMERT
    & 88.3$_{0.12}$ & 58.2$_{0.16}$ 
    & 87.4$_{0.11}$ & 60.3$_{0.16}$ 
    & 77.0$_{0.12}$ & 42.7$_{0.16}$ 
    & 84.2$_{0.13}$ & 52.0$_{0.15}$ 
    & 90.2$_{0.13}$ & 46.1$_{0.10}$ 
    & 89.1$_{0.09}$ & 35.8$_{0.13}$
    & 67.6 \\
    LyraMERT
    & 86.7$_{0.07}$ & 56.8$_{0.13}$ 
    & 85.9$_{0.08}$ & 57.4$_{0.13}$ 
    & 76.4$_{0.09}$ & 40.1$_{0.13}$ 
    & 85.0$_{0.11}$ & 53.5$_{0.14}$ 
    & 90.0$_{0.08}$ & 46.0$_{0.16}$ 
    & 88.9$_{0.05}$ & 35.1$_{0.14}$
    & 66.8 \\
    \midrule
    \textbf{CultureMERT}
    & \textbf{89.6}$_{0.09}$ & 60.6$_{0.21}$ 
    & \textbf{88.2}$_{0.20}$ & \textbf{63.5}$_{0.24}$ 
    & \textbf{79.2}$_{0.18}$ & 43.1$_{0.22}$ 
    & 86.9$_{0.10}$ & 56.7$_{0.20}$ 
    & 90.7$_{0.09}$ & 48.1$_{0.13}$ 
    & 89.4$_{0.09}$ & 35.9$_{0.16}$
    & \textbf{69.3} \\
    CultureMERT-TA
    & 89.0$_{0.12}$ & \textbf{61.0}$_{0.18}$ 
    & 87.5$_{0.10}$ & 59.3$_{0.13}$ 
    & 79.1$_{0.11}$ & 43.3$_{0.13}$ 
    & \textbf{87.3}$_{0.08}$ & \textbf{57.3}$_{0.19}$ 
    & 90.8$_{0.06}$ & 49.1$_{0.15}$
    & 89.6$_{0.10}$ & 36.4$_{0.14}$
    & 69.1 \\
    \midrule
    (Previous) SOTA
    & 87.7 \cite{DBLP:conf/ismir/PapaioannouBP23} & 57.7 \cite{DBLP:conf/ismir/PapaioannouBP23}
    & 86.5 \cite{DBLP:conf/ismir/PapaioannouBP23} & 63.1 \cite{DBLP:conf/ismir/PapaioannouBP23} & 77.0 \cite{DBLP:conf/ismir/PapaioannouBP23} & 43.9 \cite{DBLP:conf/ismir/PapaioannouBP23} & 85.4 \cite{DBLP:conf/ismir/PapaioannouBP23} & 54.3 \cite{DBLP:conf/ismir/PapaioannouBP23} & \textbf{92.4} \cite{DBLP:conf/ismir/PapaioannouBP23} & \textbf{53.7} \cite{DBLP:conf/ismir/PapaioannouBP23} & \textbf{92.7} \cite{DBLP:conf/ismir/HuangJLGLE22} & \textbf{41.4} \cite{DBLP:conf/ismir/CastellonDL21} & - \\
    \bottomrule
  \end{tabular}
  \caption{\textbf{Evaluation Results (ROC-AUC and AP) of Pre-Trained and Culturally Adapted MERT Models on Diverse Music Auto-Tagging Tasks.} We report averages across five random seeds with standard deviations as subscripts. The "Avg." column represents the average performance across all datasets and evaluation metrics for each model. The results highlight the impact of multi-cultural CPT and model merging via task arithmetic on cross-cultural adaptation and transfer.}
  \label{tab:evaluation_results}
\end{table*}

\noindent
\textbf{Multi-Cultural CPT}
In Stage 1, training runs for 2,250 steps with a 10\% linear warm-up period, using 100 hours of the dataset.
Optimization follows AdamW \cite{DBLP:conf/iclr/LoshchilovH19} with $\beta_1=0.9$, $\beta_2=0.999$, and $\epsilon=1e^{-5}$. Training employs an effective batch size of 32 recordings (160 seconds), with gradient accumulation over 8 steps. The maximum learning rate is set to $\eta_{\text{max}}=5e^{-4}$, followed by a cosine decay to a minimum of $\eta_{\text{min}}=5e^{-5}$. Gradient clipping is applied with a norm of $1.0$ to prevent exploding gradients. 
In Stage 2, training extends to 14,625 steps with a 1\% warm-up period, using the full 650-hour dataset.
Optimization follows AdamW with $\beta_1=0.9$, $\beta_2=0.95$, and $\epsilon=1e^{-5}$, maintaining the same batch size as Stage 1. The learning rate decays from a maximum value of $\eta_{\text{max}}=5e^{-5}$ to $\eta_{\text{min}}=5e^{-6}$. Gradient clipping remains at $1.0$. 

\noindent
\textbf{Single-Culture CPT}
In Stage 1, we train on 60 hours
for a total of 1,350 training steps. In Stage 2, we expand training to the full 200-hour dataset for 4,500 steps.
We employ the same optimizers, batch size, and learning rate schedules as in the multi-cultural CPT. For Lyra, due to its smaller size (50 hours), we train on 20 hours in Stage 1 (450 steps) and then on the full dataset in Stage 2 (1,125 steps).

\begin{figure}[t]
  \centering
  \includegraphics[width=0.47\textwidth]{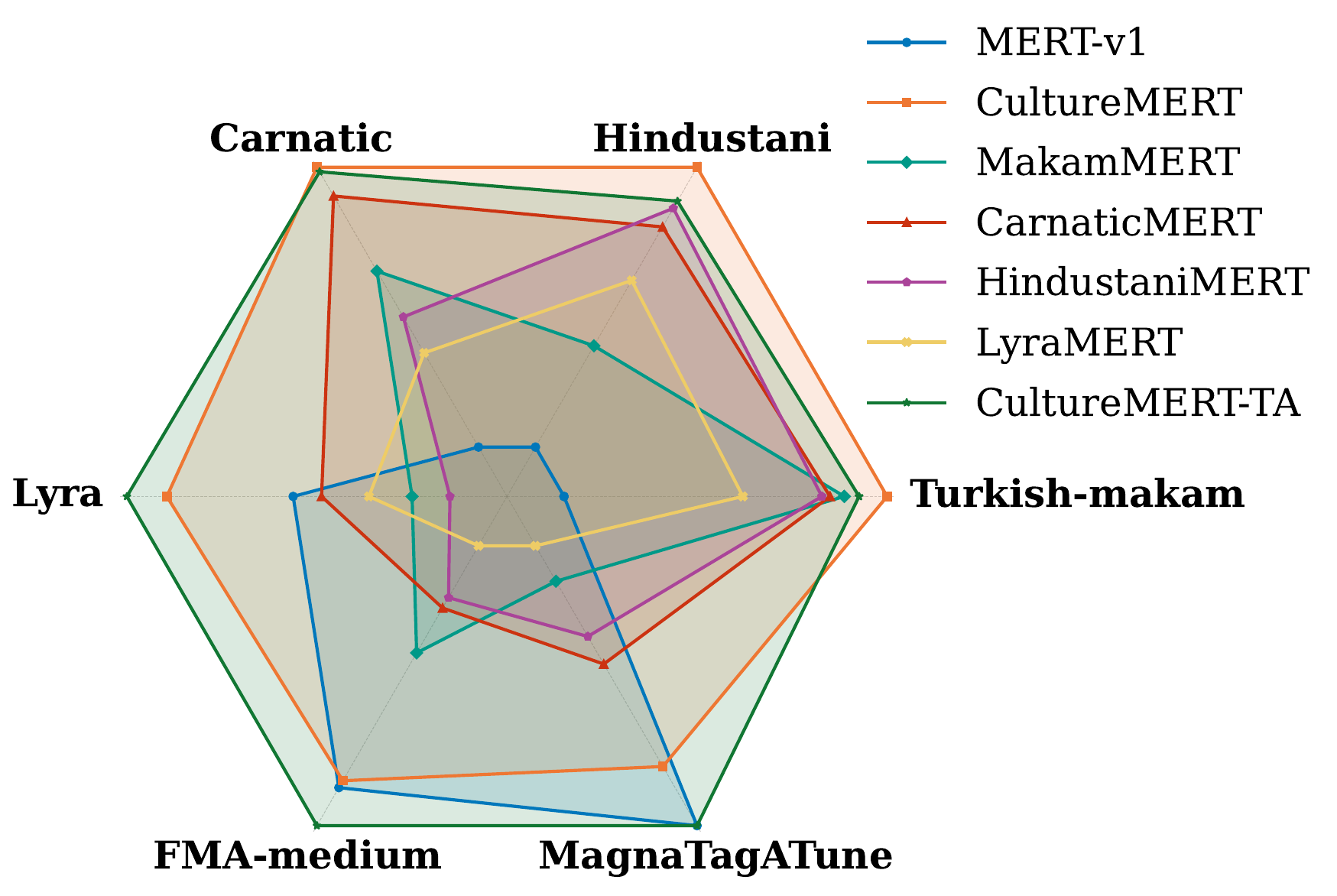}
  \caption{\textbf{Cross-Cultural Transferability.} Relative ROC-AUC performance across datasets, highlighting key trends in cross-cultural transfer. \texttt{CultureMERT} generalizes well to non-Western datasets, while task arithmetic performs on par in these settings and even surpasses both the pre-trained and multi-culturally adapted models on Western benchmarks (FMA-medium, MTAT) and Lyra.}
  \label{fig:cross_cultural_transfer_radar}
\end{figure}

\section{Results and Discussion}\label{sec:results_discussion}

As shown in Table~\ref{tab:evaluation_results}, \texttt{CultureMERT}, adapted via multi-cultural continual pre-training, consistently outperforms the original \texttt{MERT-v1} model across all non-Western tasks and evaluation metrics, achieving an average improvement of 4.9\%. It also surpasses the single-culture adapted models on average, suggesting that incorporating culturally diverse data during CPT benefits all non-Western traditions by improving the quality of representations for each individual culture, thereby enhancing generalization. Notably, \texttt{CultureMERT} achieves this with minimal forgetting on Western benchmarks (0.05\% average drop across ROC-AUC and AP), demonstrating the efficacy of our approach.
We further observe that single-culture adapted models tend to perform best on their respective in-domain tasks for well-resourced traditions, reaffirming the effectiveness of CPT for domain-specific adaptation \cite{DBLP:conf/acl/GururanganMSLBD20}. However, even low-resource adaptation, as in the case of \texttt{LyraMERT} trained on just 50 hours, leads to noticeable gains across other non-Western tasks, indicating that even limited cultural exposure can significantly boost cross-cultural generalization.
Moreover, task arithmetic performs comparably to \texttt{CultureMERT} on non-Western tasks and even surpasses it on Western benchmarks and Lyra, demonstrating that weight-space merging of culturally specialized models can serve as an effective, training-free alternative to multi-cultural CPT—provided such models are available. Interestingly, it also outperforms the unadapted base model by 0.4\% on average across Western tasks.
Notably, only the multi-cultural models, \texttt{CultureMERT} and \texttt{CultureMERT-TA}, outperform \texttt{MERT-v1} on Lyra, where the latter already serves as a strong baseline. This further underscores the effectiveness of multi-cultural adaptation, particularly in low-resource and transfer settings.
Finally, \texttt{CultureMERT} and \texttt{CultureMERT-TA} surpass previous state-of-the-art (SOTA) results on all non-Western music tagging tasks, with the best task arithmetic variant obtained using $\lambda = 0.2$ (see Figure~\ref{fig:task_arithmetic_results_per_dataset}).

\subsection{Cross-Cultural Transfer}\label{subsec:cross_cultural_results}

As illustrated in Figure~\ref{fig:cross_cultural_transfer_radar}, continual pre-training on one musical tradition can benefit others to varying degrees, revealing asymmetries in cross-cultural transfer effectiveness. For instance, we observe strong transfer between Turkish-makam and Carnatic music, with models adapted to either tradition generalizing well to the other. This aligns with their shared theoretical foundations as modal frameworks that emphasize microtonality and improvisation, serving similar roles in their respective cultures \cite{karomat200612}.
Additionally, the strong performance of the Carnatic-adapted model on the Hindustani domain reinforces the musical proximity between these traditions, particularly in their shared use of \textit{raga} (melodic mode) and \textit{tala} (rhythmic framework) \cite{DBLP:conf/ismir/KoduriMSS11}.
Interestingly, the model adapted to Carnatic music appears to be the most consistently transferable among single-culture adaptations, achieving strong results not only within Indian classical traditions but also generalizing well to Turkish-makam and Lyra.

\subsection{Token-Level Culture Similarity}\label{subsec:token_culture_similarity}

\begin{figure}[ht]
    \hspace{0.05em}
    \scalebox{0.9}
    {\includegraphics[width=0.51\textwidth]{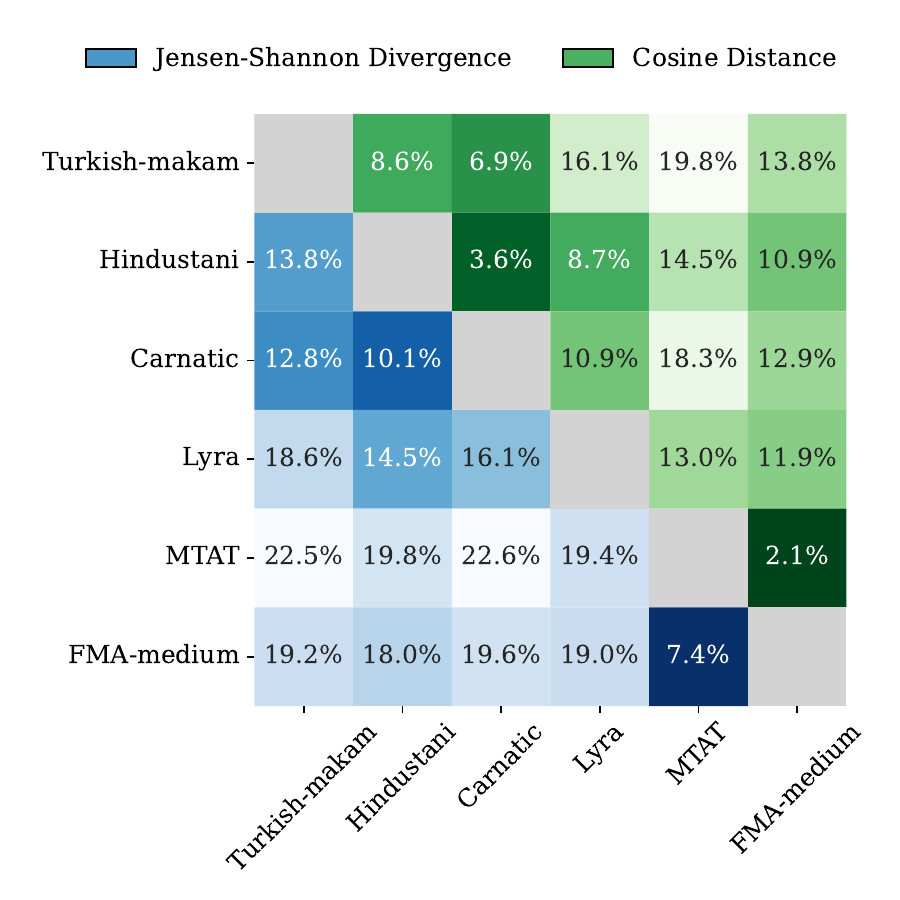}}
    \caption{\textbf{Token Similarity Across Cultures.} Pairwise similarity between acoustic token distributions extracted from the EnCodec NAC model \cite{DBLP:journals/tmlr/DefossezCSA23}. Similarity scores are averaged across $8$ codebooks, each containing $1024$ discrete codewords (acoustic pseudo-tokens).}
    \label{fig:token_jsd_cosine_heatmap}
\end{figure}

To further examine cross-cultural similarities in our data, we analyze token overlap across musical traditions using both the Jensen-Shannon divergence (JSD) and cosine distance between token distributions extracted from the EnCodec model \cite{DBLP:journals/tmlr/DefossezCSA23}, which serves as our audio tokenizer. 
Lower values in both metrics indicate greater similarity. 
Our analysis, as shown in Figure~\ref{fig:token_jsd_cosine_heatmap}, reveals strong token-level similarity among non-Western traditions, particularly between Hindustani and Carnatic music. In contrast, Western datasets (MTAT, FMA-medium) are highly similar to each other but notably dissimilar from non-Western traditions. Greek traditional music (Lyra), while distinct, aligns more closely with non-Western traditions than Western ones.
Interestingly, these findings correlate with our results on cross-cultural transfer (Section~\ref{subsec:cross_cultural_results}), suggesting that token-level similarity metrics can serve as predictors of positive cross-cultural transfer.
This insight has practical implications: such similarity metrics can guide the selection and refinement of pre-training data mixtures during CPT, or inform the adjustment of arithmetic operations when merging models via task arithmetic.
Similar approaches for quantifying language similarity and predicting positive cross-lingual transfer, based on the similarity of extracted linguistic or acoustic tokens, have been explored in both the text \cite{DBLP:conf/tsd/GogoulouLBN24, DBLP:journals/corr/abs-2501-14491} and speech domains \cite{DBLP:conf/sigtype/SanPAHKAJ24}.

\subsection{Task Arithmetic Scaling Factor}\label{subsec:ta_results_factor}

\begin{figure}[ht]
  \centering
  \includegraphics[width=0.47\textwidth]{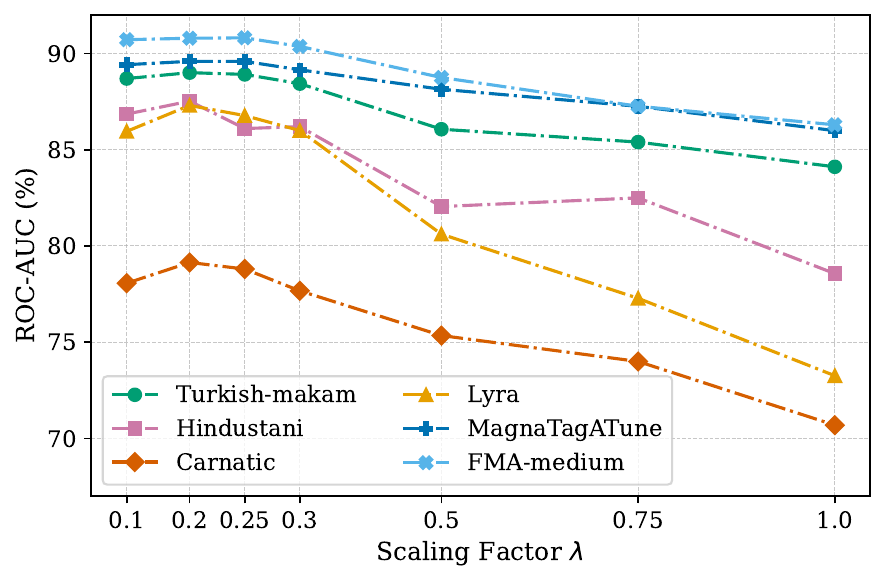}
  \caption{\textbf{Effect of Scaling Factor $\lambda$ on Task Arithmetic Performance.} The ROC-AUC scores across six diverse music tagging tasks demonstrate how varying $\lambda$ impacts task arithmetic when merging the four non-Western single-culture adapted models.}
  \label{fig:task_arithmetic_results_per_dataset}
\end{figure}

A key consideration in task arithmetic is the choice of the scaling factor $\lambda$, which controls the balance between task vectors. Prior work \cite{DBLP:conf/iclr/YangW00G0T24, DBLP:conf/eacl/ParovicVK24} has shown that suboptimal values can significantly degrade performance in multi-task model merging.
We systematically evaluate different values of a shared scaling factor $\lambda \in \{0.1, 0.2, 0.25, 0.3, 0.5, 0.75, 1.0\}$, applied uniformly across all task vectors, including the special case of weight averaging ($\lambda = 0.25$). 
We similarly observe a consistent trend: ill-suited values, such as $\lambda = 1.0$, result in poor performance across all benchmarks, as shown in Figure~\ref{fig:task_arithmetic_results_per_dataset}.

\section{Conclusions}\label{sec:conclusions}

In this paper, we introduce \texttt{CultureMERT-95M}, a multi-culturally adapted music foundation model developed via continual pre-training on diverse non-Western musical traditions. We propose a two-stage CPT strategy that incorporates learning rate re-warming and staged adaptation for stable training. Cross-cultural evaluation demonstrates that \texttt{CultureMERT-95M} consistently outperforms the base \texttt{MERT-v1-95M} model on non-Western music tagging tasks, surpassing prior state-of-the-art methods while preserving performance on Western benchmarks. Additionally, we investigate task arithmetic, which offers a strong alternative to multi-cultural CPT by effectively merging culturally specialized models in weight space.

While our results are promising, several limitations remain. The frozen EnCodec tokenizer used in the MERT architecture may be suboptimal for encoding culturally diverse musical languages, as it was pre-trained on Western music. Future directions include scaling to additional musical cultures, exploring alternative architectures, extending evaluation beyond sequence-level classification tasks, conducting fine-grained ablation studies, and investigating whether the proposed two-stage CPT strategy remains necessary under less constrained computational budgets.

\section{Ethics Statement}\label{sec:ethics_statement}

\subsection{Cultural Framing and Interpretive Scope}\label{subsec:cultural_interpretation}
We acknowledge the limitations of framing music within a "Western" versus "non-Western" dichotomy. While such terminology is commonly used in computational research for convenience, it risks oversimplifying the rich diversity of global musical traditions. 
Furthermore, this work does not aim to establish or analyze cross-cultural similarities from an ethnomusicological perspective. Our analysis of cross-cultural transferability should be considered in light of potential limitations in the representativeness and coverage of the corpora used.

\subsection{Responsible Use}\label{subsec:responsible_use}
Careful consideration is advised before deploying these models in real-world contexts, as they may still reflect cultural and dataset-specific biases.
Some of the datasets used in this work are not publicly available and were obtained under research-use agreements. The released models should not be used in commercial or generative applications without explicit attention to cultural representation, appropriate licensing, and the consent of the relevant communities or dataset curators.

\section{Acknowledgments}\label{sec:acknowledgments}

We would like to thank the reviewers and Georgios Paraskevopoulos for their valuable and constructive feedback, which helped us improve this work. We also gratefully acknowledge the Music Technology Group (MTG) at Universitat Pompeu Fabra for providing access to datasets used in this study.
This work has been partially supported by project MIS 5154714 of the National Recovery and Resilience Plan Greece 2.0 funded by the European Union under the NextGenerationEU Program.

\bibliography{ISMIRreferences}

%
%
%
%

\end{document}